\documentclass{article}
\usepackage{spconf,amsmath,graphicx}
\usepackage{amssymb,amsmath,bm}
\usepackage{graphicx}
\usepackage{multirow}
\usepackage{enumitem}
\usepackage{subfig}
\graphicspath{{./pdf/}}
\DeclareGraphicsExtensions{.pdf}

\newcommand{\tabincell}[2]{\begin{tabular}{@{}#1@{}}#2\end{tabular}}

\def\vec#1{\ensuremath{\bm{{#1}}}}
\def\mat#1{\vec{#1}}

\title{Comparison of Multiple Features and Modeling Methods for Text-dependent Speaker Verification}
\name{Yi Liu$^1$, Liang He$^1$, Yao Tian$^1$, Zhuzi Chen$^1$, Jia Liu$^1$, Michael T. Johnson$^2$ \thanks{The work is supported by National Natural Science Foundation of China under Grant No. 61370034, No. 61403224 and No. 61273268.}}
\address{$^1$Tsinghua National Laboratory for Information Science and Technology, \\
Department of Electronic Engineering, Tsinghua University, Beijing 100084, China \\
$^2$Department of Electrical and Computer Engineering, University of Kentucky \\
453 F. Paul Anderson Tower, Lexington, KY 40506-0046}

\begin{document}
\ninept
\maketitle
\begin{abstract}
  Text-dependent speaker verification is becoming popular in the speaker recognition society. However, the conventional i-vector framework which has been successful for speaker identification and other similar tasks works relatively poorly in this task. Researchers have proposed several new methods to improve performance, but it is still unclear that which model is the best choice, especially when the pass-phrases are prompted during enrollment and test. In this paper, we introduce four modeling methods and compare their performance on the newly published RedDots dataset. To further explore the influence of different frame alignments, Viterbi and forward-backward algorithms are both used in the HMM-based models. Several bottleneck features are also investigated. Our experiments show that, by explicitly modeling the lexical content, the HMM-based modeling achieves good results in the fixed-phrase condition. In the prompted-phrase condition, GMM-HMM and i-vector/HMM are not as successful. In both conditions, the forward-backward algorithm brings more benefits to the i-vector/HMM system. Additionally, we also find that even though bottleneck features perform well for text-independent speaker verification, they do not outperform MFCCs on the most challenging Imposter-Correct trials on RedDots.


\end{abstract}
\section{Introduction}
\label{sec:intro}
During the past two decades, automatic speaker verification (ASV) has made a great deal of progress for many practical applications. Generally speaking, speaker verification can be categorized into two groups: \emph{text-independent} and \emph{text-dependent}. Unlike text-independent ASV, speaker and pass-phrase identities are both taken into consideration in text-dependent speaker verification. Thus the trials in text-dependent ASV consist of: (1) target speaker produces correct content, (2) target speaker produces wrong content, (3) imposter speaker produces correct content, and (4) imposter speaker produces wrong content. Only the first target-correct combination is accepted by the text-dependent system.

While text-independent ASV is much more flexible and achieves high accuracy for long records, the performance dramatically degrades when the utterance is too short \cite{ti_short}. Text-dependent ASV, in contrast, provides acceptable performance in this scenario if pass-phrases are properly designed \cite{rsr2015}. This is important for many commercial applications, e.g. user authorization, because the duration of user speech keeps just a few seconds. Currently, text-dependent ASV is becoming much more popular and attracting the attention of institutes and companies \cite{utt_verification, end2end_microsoft}.

Researchers have proposed several techniques to tackle text-dependent ASV. There are two main categories of modeling: template-based and statistical-based. The template-based modeling utilizes dynamic time warping (DTW) to calculate the distance between enrollment and test speech. Two utterances are assigned to the same speaker and pass-phrase if the distance is short enough. The feature vectors in the template-based modeling are required to reflect the speaker characteristics and the lexical content simultaneously. Posteriorgrams based on speaker-specific Gaussian mixture model (GMM) were first used as features in \cite{gauss_post_dtw}. Later, deep neural networks (DNNs) were introduced, with DNN posteriors chosen to form the posteriorgrams instead \cite{nn_post_dtw}. Although the DNN approach gives more meaningful phonetic posteriors than GMM, it cannot discriminate different speakers. DNN-based online i-vectors were presented as features in \cite{ivec_dtw} to solve this problem. Another important factor in DTW is the distance metric. Kullback-Leibler (KL) divergence is a natural choice to measure the distance between posteriorgrams, while for online i-vectors, cosine distance is more efficient. The idea of template-based modeling is quite straightforward, but only works for the fixed-phrase condition.

The statistical-based modeling approaches originate from traditional text-independent ASV. Driven by NIST speaker recognition evaluation (SRE), text-independent speaker verification technologies have developed rapidly \cite{ti_short, ti_overview_2000}. GMM-universal background model (GMM-UBM) \cite{tandem_gmm_td}, GMM-support vector machine (GMM-SVM) \cite{gmm_svm}, and more advanced models with channel compensation, like joint factor analysis (JFA) \cite{jfa_digit} and i-vector framework\cite{rsr2015} have all been applied to text-dependent ASV. To explicitly capture sequence information, the hidden Markov model (HMM) is leveraged across these models. The HMM can be built empirically with no knowledge about the actual content \cite{fix_phrase_hmm1, fix_phrase_hmm2}, or composed according to the text transcriptions \cite{gmm_hmm, ivec_hmm_result}. The HMM first aligns each frame to a certain state, and then further computations, such as speaker adaptation, likelihood calculation, and sufficient statistics extraction, are based on the alignment result.

Although i-vector modeling has become dominant in text-independent speaker verification, for text-dependent verification, the former GMM-UBM still performs better \cite{rsr2015}. Efforts have been made to improve i-vector performance, but a lack of common datasets and evaluation protocols has made it difficult to evaluate and compare their results. It is somewhat unclear which method is the best choice for text-dependent speaker verification. Just recently some text-dependent speaker verification datasets have been made available. In this paper, we choose several state-of-the-art modeling methods and compare their performance under both fixed- and prompted-phrase condition on the RedDots dataset \cite{reddots}. Because the alignment process is crucial for speaker verification, we further compare the results between Viterbi and forward-backward (FB) algorithms for alignment and check the influence of varying alignment granularity.

Apart from modeling methods, we also investigate multiple bottleneck (BN) features in text-dependent ASV. Such BN features are extracted from a narrow hidden layer of a DNN and can be easily coupled with other models.
The DNN can be trained in different styles, leading to two major groups of BN features. The first one is the utterance-level BN feature which is also called as \emph{speaker embedding}. Speaker embedding has the ability to discover speaker characteristics and encode the entire utterance into a fixed-length vector. To extract speaker embedding, the neural network is often trained in end-to-end style to ensure the speaker information can be effectively involved. This idea was first proposed by Google \cite{end2end_google} and was then further developed by \cite{end2end_microsoft, end2end_nin, deep_speaker}. Unlike speaker embedding for the entire utterance, frame-level BN features are extracted frame-by-frame. In this case, the network is often trained to classify phrases, phonemes, speaker identities or their combination \cite{deep_feature_sjtu, bn_feature_crim}. Different BN features contain different information about the target tasks. Frame-level BN features can be used as conventional cepstral features.

Even though speaker embedding achieves promising performance in text-dependent speaker verification, it requires huge amounts of in-domain data to train a robust model \cite{bn_feature_crim}. This constraint may not be satisfied in some conditions. Therefore, frame-level BN features still play an important role in speaker verification. Despite the success in text-independent speaker verification, the effectiveness of various BN features in text-dependent ASV is still under exploration. In this paper, we investigate the performance of BN features extracted from phonetic and speaker discriminant DNNs.

Another recently published paper \cite{ivector_hmm} has also compared several modeling methods and their combination with BN features in text-dependent speaker verification.
Differences between that work and the results presented here include:
\begin{itemize}
 \item The systems in our paper are evaluated under both fixed- and prompted-phrase condition. In prompted-phrase verification, phoneme mismatch exists between enrollment and test utterances, which may have different impacts to these modeling methods. We should also notice that the prompted-phrase condition is more attractive in commercial verification applications.
 \item Two different HMM alignment methods, i.e. Viterbi and forward-backward algorithms, are compared in this paper.
 \item The BN features are not only extracted from a phonetic-dependent DNN, but also from a speaker discriminant DNN.
\end{itemize}

The remainder of this paper is organized as follows. The various modeling methods and different alignment schemes are introduced in Section 2. Section 3 describes different BN features used in this paper. Experimental setup and results are presented in Section 4 and Section 5. Finally, Section 6 concludes the paper.

\section{Modeling methods for text-dependent verification}
\label{sec:modeling}
In this section, we present four modeling methods most frequently used in text-dependent ASV, as well as two HMM alignment algorithms.

\subsection{GMM-UBM}
\label{ssec:gmm_ubm}
For completeness, we first introduce the classical GMM-UBM which is still the baseline in text-dependent ASV.

The concept of GMM-UBM has long been proposed in text-independent speaker verification \cite{gmm_ubm}. After training the UBM model, the speaker model is adapted using maximum a posteriori (MAP) estimation. Given the observation $\vec{x}_t$, the posterior of component $c$ (referred as \emph{frame alignment}) is:
\begin{equation}\label{posterior_probability}
  P(c|\vec{x}_t) = \frac{w_c \mathcal{N}(\vec{x}_t|\vec{\mu}_c,\mat{\Sigma}_c)}{\sum_{m} w_m \mathcal{N}(\vec{x}_t|\vec{\mu}_m,\mat{\Sigma}_m)}
\end{equation}
where $w_c$, $\vec{\mu}_c$ and $\mat{\Sigma}_c$ denote the weight, mean and covariance of component $c$, respectively. The zero- and first-order sufficient statistics are accumulated over the utterances of speaker $\mathcal{S}$:
\begin{equation}\label{zero_order}
  N_c = \sum_{t} P(c|\vec{x}_t)
\end{equation}
\begin{equation}\label{first_order}
  \vec{F}_c = \sum_{t} P(c|\vec{x}_t)(\vec{x}_t-\vec{\mu}_c)
\end{equation}
The mean of every component is updated with the statistics while other parameters are kept unchanged:
\begin{equation}\label{adapt}
  \hat{\vec{\mu}}_c = \alpha_c \cdot \vec{F}_c + \vec{\mu}_c
\end{equation}
where $\alpha_c = 1 / (N_c+r)$, and $r$ is the relevant factor. During the test phase, the log-likelihood ratio is computed against two models:
\begin{equation}\label{llr}
  \text{LLR} = \sum_t \log p(\vec{x}_t|\Lambda^{\text{spkr}}) - \log p(\vec{x}_t | \Lambda^{\text{ubm}})
\end{equation}
where $\Lambda^{\text{spkr}}$ and $\Lambda^{\text{ubm}}$ are the speaker model and UBM, respectively.

\subsection{i-vector modeling}
\label{ssec:ivector}
Session variability between enrollment and test speech is a significant issue in speaker verification. The nuisance variability is explicitly modeled in i-vector framework \cite{ivector}. Motivated by GMM-UBM, it is hypothesized speaker information can be effectively coded by the GMM mean vectors. The mean vectors $\vec{\mu}_c$ are concatenated into a super-vector and the i-vector modeling is expressed as:
\begin{equation}\label{ivector_modeling}
  \vec{M}=\vec{m}+\mat{T}\vec{w}
\end{equation}
where $\vec{m}=[\vec{\mu}'_1,\ldots,\vec{\mu}'_C]'$ is the UBM super-vector, $\vec{M}$ is the utterance GMM super-vector, $\mat{T}$ is a low rank matrix and $\vec{w} \sim \mathcal{N}(\vec{0}, \mat{I})$ is a random variable. The matrix $\mat{T}$ is estimated first and the i-vector for utterance $s$ is the MAP point estimation of $\vec{w}$ which is denoted as:
\begin{equation}\label{ivector}
  E[\vec{w}(s)] = \mat{L}^{-1}(s)\mat{T}'\mat{\Sigma}^{-1}\vec{F}(s)
\end{equation}
where $\mat{L}(s)$ is the precision matrix of the posterior distribution of $\vec{w}(s)$:
\begin{equation}\label{ivector_prec}
  \mat{L}(s)= \mat{I} + \mat{T}'\mat{\Sigma}^{-1}\mat{N}(s)\mat{T}
\end{equation}
where $\mat{\Sigma}$ and $\mat{N}(s)$ are block diagonal matrices whose diagonal blocks are $\mat{\Sigma}_c$ and $N_c(s)\mat{I}$, and $\vec{F}(s)=[\vec{F}'_1(s),\ldots,\vec{F}'_C(s)]'$ is a super-vector of the first-order statistics.

\subsection{GMM-HMM}
\label{ssec:gmm_hmm}
When calculating Eq. \ref{posterior_probability} in GMM-UBM, the underlying hypothesis is that every component of the UBM denotes a pronunciation attribute and the posterior measures how this frame aligns to the attribute. The attributes are formed using unsupervised clustering in GMM-UBM which leads to a suboptimal alignment result. In text-dependent ASV, however, the situation changes. Given transcriptions, there is an opportunity to explicitly align frames to more meaningful phonemes rather than unsupervised clustered attributes.

In GMM-HMM modeling, a phoneme recognizer is first trained using the standard ASR recipe. We use 3-state mono-phone HMMs to construct this recognizer. The emission probability of every state is modeled by a GMM. Say we have $F$ mono-phones ($F=39$ in our English lexicon), and a $G$-mixture GMM for each mono-phone state. Thus the GMM-HMM has $3FG$ mixture components in total. Compared to GMM-UBM, these mixtures are better separated in the phoneme space.

Given a transcription, a graph of HMM is composed. An algorithm can be applied to calculate the posterior $P((j,g)|x_t)$ of $\vec{x}_t$ occupying the $g$-th mixture of state $j$. For simplicity, We will denote it as $P_t(j,g)$ in the following.

The Viterbi and forward-backward (FB) algorithms are two means to align frames to states and mixtures. The Viterbi algorithm finds the 1-best path where frame $\vec{x}_t$ is aligned to the most possible state $q_t$, i.e., $P(j|\vec{x}_t)=0$ if $j \neq q_t$. In this case, the posterior $P_t(j, g)$ can be derived by Bayes formula:
\begin{equation}\label{align_posterior}
\begin{split}
  P_t(j,g) &= P(j|\vec{x}_t) \cdot P(g|j,\vec{x}_t)\\
  &=  P(j|\vec{x}_t) \cdot
   \frac{w_{(j,g)} \mathcal{N}\left(\vec{x}_t|\vec{\mu}_{(j,g)},\mat{\Sigma}_{(j,g)}\right)}
   {\sum_{m=1}^{G} w_{(j,m)} \mathcal{N}\left(\vec{x}_t|\vec{\mu}_{(j,m)},\mat{\Sigma}_{(j,m)}\right)}
\end{split}
\end{equation}
Speaker adaption is the same with Eq. \ref{adapt} except mixtures here are phonetic dependent. During the test phase, the Viterbi-based log-likelihood ratio is expressed as:
\begin{equation}\label{align_llr_vit}
  \text{LLR}_{\text{Vit}} = \sum_t \log p(\vec{x}_t | \Lambda^{\text{spkr}}_{q_t}) - \log p(\vec{x}_t | \Lambda^{\text{ubm}}_{q_t})
\end{equation}
where $\Lambda^{\text{spkr}}_{q_t}$ and $\Lambda^{\text{ubm}}_{q_t}$ are the speaker and background models of state $q_t$.

Unlike the Viterbi algorithm, the FB algorithm directly computes the probability $P_t(j, g)$ and the hard boundary between states is replaced by a soft boundary function which is computed from the forward and backward probabilities \cite{rabiner}. Hence, each frame may have non-zero posteriors for all states. The log-likelihood then becomes:
\begin{equation}\label{align_llr_fb}
\begin{split}
  \text{LLR}_{\text{FB}} = \sum_t \Big( &\log \sum_{j,g}P_t(j,g) \mathcal{N}\left(\vec{x}_t | \hat{\vec{\mu}}_{(j,g)}, \mat{\Sigma}_{(j,g)}\right) \\
   - &\log \sum_{j,g} P_t(j,g) \mathcal{N}\left(\vec{x}_t | \vec{\mu}_{(j,g)}, \mat{\Sigma}_{(j,g)}\right) \Big)
\end{split}
\end{equation}
where $\hat{\vec{\mu}}_{(j,g)}$ and $\vec{\mu}_{(j,g)}$ are the adapted and background means of state $j$ and mixture $g$.

We would expect the FB algorithm to perform better because it gives alignment with finer granularity, especially at the regions between phonemes. We should also notice that the FB algorithm requires more computation than the Viterbi alignment. If $T$ denotes the number of input vectors and $N$ is the state number in the composite HMM, the time complexity of both algorithms is $\mathcal{O}(N^2 T)$. However, the FB algorithm needs about twice the number of computations because of the additional backward process \cite{rabiner}. 


\subsection{i-vector/HMM}
\label{ssec:ivector_hmm}
Eq. \ref{ivector} and Eq. \ref{ivector_prec} indicate that the sufficient statistics, and thus the posteriors $P(c|\vec{x}_t)$, are critical to train and compute i-vectors. In text-independent i-vector framework, the posteriors are calculated by GMM, or senone-based DNN \cite{dnn_ivector}. This can be extended to the text-dependent case where the HMM Viterbi or FB alignment is used: 
\begin{equation}\label{align_zero_order}
  N_{(j,g)} = \sum_{t} P_t(j,g)
\end{equation}
\begin{equation}\label{align_first_order}
  \vec{F}_{(j,g)} = \sum_{t} P_t(j,g)\left(\vec{x}_t-\vec{\mu}_{(j,g)}\right)
\end{equation}

Because the duration is short in text-dependent ASV, the phoneme contents of a single utterance are always limited. The HMM alignment of each frame tends to concentrate on the correct phoneme states, making many entries of the statistics zeros. This sparseness helps to reduce the uncertainty of the i-vector and is useful to reject wrong phrases in the fixed-phrase condition \cite{ivector_hmm}.
Only cosine distance is used for i-vector scoring in this paper because intersession compensation techniques are ineffective if the phrases in the evaluation set differ from the background training data \cite{ivec_hmm_result}.

\begin{figure}[tb]
    \centering
    \includegraphics[width=0.8\linewidth]{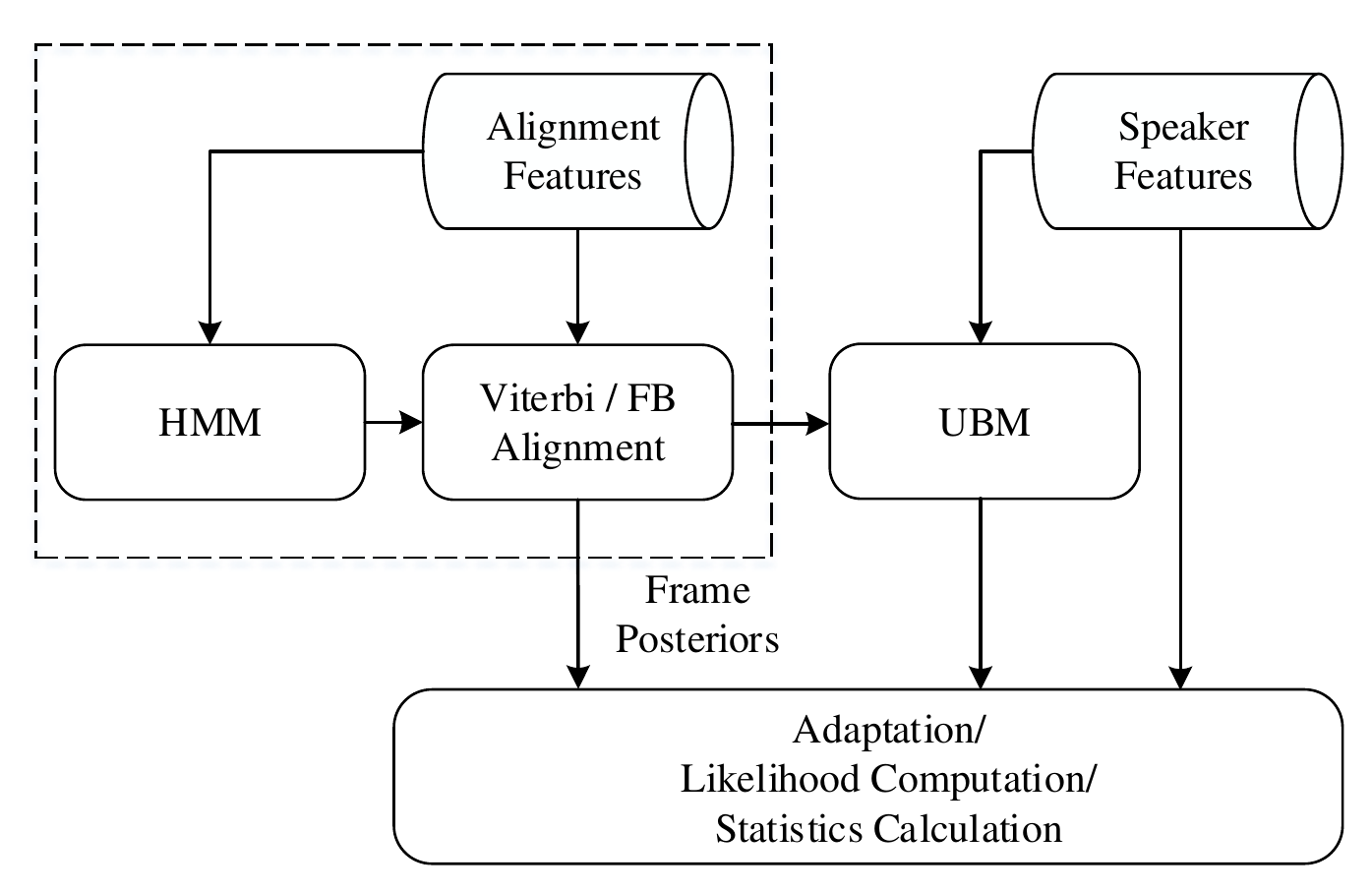}
    \caption{{ The flow diagrams of the HMM-based models. The alignment features are used for the HMM training and alignment while the speaker features are used for GMM and i-vector modeling.}}
    \label{fig:two_model}
\end{figure}

It is worth mentioning that, similar to the two-model approach in text-independent speaker verification \cite{bn_feature_overview}, features used for HMM training and alignment may differ from the ones used for GMM or i-vector modeling. This is illustrated in Fig. \ref{fig:two_model}. After calculating the posteriors, an auxiliary UBM is re-estimated for every state using speaker features. Speaker adaptation, likelihood computation and statistics calculation are based on these new GMMs. In this paper, we keep the alignment feature unchanged while using different speaker features for GMM-HMM and i-vector/HMM modeling.

\section{Bottleneck features}
\label{sec:feature}
Bottleneck features have been successfully applied in speech recognition and text-independent speaker verification \cite{bn_feature_overview}. Classical BN features are extracted from a DNN trained for phoneme classification. A hidden layer in the DNN has significantly fewer nodes than its surrounding layers. The BN feature is the output of this bottleneck layer which filters the information flowing through and retains the most useful part. In text-independent speaker verification, the mixture components of the GMM trained using the phonetic-dependent BN features are more connected with the lexical content. This explicit connection makes the GMM alignment (Eq. \ref{posterior_probability}) more accurate, thus improves the final verification performance.

In text-dependent speaker verification, however, the alignment is generated by the HMM Viterbi or FB algorithms. This fact seems to limit the effectiveness of the phonetic BN features. Moreover, the BN feature is discriminative against phonetic content and may eliminate the inter-speaker variability. This is, in turn, detrimental for speaker verification. 

Apart from phonetic DNNs, speaker discriminant DNNs have been a recent trend \cite{deep_speaker_feature}. Speaker discriminant DNNs have the same structure with phonetic DNNs but replace the output labels with speaker identities. The BN features extracted from this network do not provide better frame alignments but attempt to directly separate speakers in the feature space. The features can be simply averaged across an utterance to get a speaker embedding, called \emph{d-vector}. When the data size and the number of speakers become large enough, d-vector outperforms i-vector for short utterance \cite{deep_speaker_feature_new}. We do not follow the d-vector recipe but use the speaker discriminant BN features as cepstral coefficients and combine them with the above modeling approaches.

\section{Experiments}
\label{sec:experiments}

\subsection{Data}
\label{ssec:data}
We choose RedDots for the evaluation. RedDots is an ongoing crowdsourcing dataset collected especially for text-dependent ASV \cite{reddots}. RedDots consists of four parts of increasing lexical variability. Part-1 to Part-3 denote standard text-dependent conditions in which the pass-phrases remain fixed in the enrollment and test speech. In these parts, a speaker with different phrases is treated as several separated models. These parts are relatively easy because the lexicon gives extra information and the intra-speaker variability is constrained. On Part-4, phrases are prompted during enrollment and test, and none of the speaker enrollment data covers all phonemes. This part is very close to the traditional text-independent task and is more difficult than the other parts. 

The RedDots dataset contains 62 speakers, including 49 males and 13 females. The experiments are conducted on the male part of Part-1 and Part-4. The female part is excluded in our experiments because there are not enough amounts of speakers in the current release. We report the results for all three error types on Part-1, i.e., Imposter-Correct (IC), Target-Wrong (TW) and Imposter-Wrong (IW). The genuine scores are from the Target-Correct trials while the non-target scores are provided by trials of these errors. On Part-4, models are enrolled using utterances with various contents. Therefore, these models do not have information of one specific sentence. Speech recognition techniques are required to detect wrong text. To focus on speaker verification, we only present the performance on the Imposter-Correct trials on Part-4. Equal error rate (EER) and minimum detection cost function from NIST SRE08 (MDCF08) and SRE10 (MDCF10) \cite{mdcf} are the evaluation metrics.

The other datasets we used in our experiments are Librispeech \cite{librispeech} and RSR2015 \cite{rsr2015}. Librispeech is a read speech dataset with around 1000 hours of English speech. RSR2015 is released by IIR in Singapore and is also designed to support the research on text-dependent speaker verification. All these datasets are sampled at 16kHz and are publicly accessible to make the results in this paper reproducible.

\subsection{Features}
\label{ssec:feature}
\begin{itemize}[leftmargin=*]
 \item \textbf{MFCC}: MFCCs are extracted from 16kHz utterance with 40 filter-banks distributed between 0 and 8kHz. Static 19-dimensional coefficients plus energy and their delta and delta-delta form a 60-dimensional vector. CMVN is applied per utterance.
 \item \textbf{FBank}: FBank computation has the same steps with MFCC extraction, except that no DCT transform is applied. The Fbank coefficient is 120 dimension ($40+\Delta+\Delta \Delta$).
 \item \textbf{BN feature}: BN features are extracted from phonetic and speaker discriminant DNNs. We denote them as \emph{pBN} and \emph{sBN}, respectively. The configuration of the DNN structure is described below.
\end{itemize}

\subsection{Models}
\label{ssec:model}

\begin{table}[tb]
\caption{Corpora used to estimate the HMM, DNN, UBM and matrix T in the experiments.}
\label{table:data_setup}
\centering
\begin{tabular}{|c|c|c|c|}
\cline{2-4}
\multicolumn{1}{c|}{} & \tabincell{c}{Librispeech\\train-clean-100} & \tabincell{c}{Librispeech\\train-clean-360} & \tabincell{c}{RSR2015\\Part-1} \\
\cline{2-4}
\hline
HMM & X &   & X \\
\hline
DNN & X & X & X \\
\hline
UBM & X &   & X  \\
\hline
T   & X &   & X  \\
\hline
\end{tabular}
\end{table}

\begin{itemize}[leftmargin=*]
 \item \textbf{DNN}: The training data is a subset from Librispeech and RSR2015, containing about 460 and 50 hours respectively.  The fully connected DNN consists of 4 hidden layers with 1200 nodes per layer, plus a BN layer with 60 nodes. The input is FBank with symmetric 5-frame expansion, resulting in 11 frames in total. The number of nodes in the output layer is determined by different tasks. For phoneme classification, we use 2142 tied triphone states and for speaker classification, 1472 speaker identities (1172 from Librispeech and 300 from RSR2015) become the corresponding outputs. BN features are the linear outputs of the last hidden layer without any activation function. 
 \item \textbf{GMM-UBM} and \textbf{i-vector}: A gender-dependent UBM with 1024 mixtures is trained. For i-vector modeling, the rank of the matrix $\mat{T}$ is 600. Log-likelihood without normalization and cosine distance is used to generate verification scores in these two systems respectively.
 \item \textbf{HMM}: To generate the alignment for the HMM-based modeling, we use MFCCs to train the HMM. 39 mono-phones plus a silence model are used, each of which contains 3 states. To model the complexity of silence, a GMM with 16 mixtures is used for every silence state, while other states are all modeled by 8 Gaussians, resulting 984 Gaussians in total. This HMM is further extended to a triphone system and remains 2142 senones. The transcriptions for DNN training is generated by the senone alignment. Only MFCCs are used for HMM training and alignment.
 \item \textbf{GMM-HMM} and \textbf{i-vector/HMM}: The GMM of every state is re-estimated using the HMM alignments and different speaker features. The total number of mixtures in our model is 984. The dimension of i-vector is again set to 600. Viterbi and FB alignments are both investigated.
\end{itemize}

\begin{table*}[htpb]
\caption{Comparison of different modeling methods on the male part of RedDots Part-1. IC, TW, IW denote three error types, i.e., Imposter-Correct, Target-Wrong and Imposter-Wrong.}
\label{table:result_part1}
\centering
\begin{tabular}{|c|c|c|c||c|c|c||c|c|c|}
\hline
\multirow{2}{*}{Modeling} & \multicolumn{3}{c||}{IC} & \multicolumn{3}{c||}{TW} & \multicolumn{3}{c|}{IW} \\
\cline{2-10}
 & EER & MDCF08 & MDCF10 & EER & MDCF08 & MDCF10 & EER & MDCF08 & MDCF10 \\
\hline
\hline
GMM-UBM               & 2.23 & 0.0116 & 0.3685 & 6.02 & 0.0278 & 0.6070 & 0.86 & 0.0033 & 0.1758 \\
\hline
i-vector              & 4.26 & 0.0193 & 0.5079 & 9.99 & 0.0468 & 0.8072 & 1.39 & 0.0065 & 0.2902 \\
\hline
\hline
GMM-HMM Viterbi       & 1.91 & 0.0080 & 0.2649 & 1.91 & \textbf{0.0072} & 0.2452 & 0.65 & 0.0016 & 0.0478 \\
\hline
GMM-HMM FB            & \textbf{1.88} & \textbf{0.0078} & 0.2483 & 1.91 & 0.0074 & 0.2677 & 0.62 & \textbf{0.0015} & 0.0718 \\
\hline
i-vector/HMM Viterbi  & 2.31 & 0.0085 & 0.2316 & 2.00 & 0.0076 & 0.1993 & 0.65 & 0.0016 & 0.0466 \\
\hline
i-vector/HMM FB       & 2.07 & 0.0081 & \textbf{0.2082} & \textbf{1.82} & 0.0075 & \textbf{0.1829} & \textbf{0.59} & \textbf{0.0015} & \textbf{0.0404} \\
\hline
\end{tabular}
\end{table*}

\begin{table}[htpb]
\caption{Comparison of different modeling methods on the male part of RedDots Part-4. We only report the results on the Imposter-Correct trials.}
\label{table:result_part4}
\centering
\begin{tabular}{|c|c|c|c|}
\hline
Modeling & EER & MDCF08 & MDCF10 \\
\hline
\hline
GMM-UBM               & \textbf{9.03} & \textbf{0.0431} & 0.9857 \\
\hline
i-vector              & 11.64 & 0.0468 & 0.9611 \\
\hline
\hline
GMM-HMM Viterbi       & 9.78 & 0.0463 & 0.9282 \\
\hline
GMM-HMM FB            & 9.79 & 0.0460 & 0.9359 \\
\hline
i-vector/HMM Viterbi  & 9.94 & 0.0446 & \textbf{0.8453} \\
\hline
i-vector/HMM FB       & 9.52 & 0.0438 & 0.8476 \\
\hline
\end{tabular}
\end{table}

\begin{table*}[htpb]
\caption{Comparison of different features on the male part of RedDots Part-1. IC, TW, IW denote three error types, i.e., Imposter-Correct, Target-Wrong and Imposter-Wrong. pBN and sBN denote features extracted from phonetic and speaker discriminant DNNs.}
\label{table:bn_part1}
\centering
\begin{tabular}{|c|c|c|c||c|c|c||c|c|c|}
\hline
\multirow {2}{*}{Modeling} & \multicolumn{3}{c||}{IC} & \multicolumn{3}{c||}{TW} & \multicolumn{3}{c|}{IW} \\
\cline{2-10}
 & EER & MDCF08 & MDCF10 & EER & MDCF08 & MDCF10 & EER & MDCF08 & MDCF10 \\
\hline
\hline
MFCC         & \textbf{1.91} & \textbf{0.0080} & \textbf{0.2649} & 1.91 & 0.0072 & 0.2452 & 0.65 & \textbf{0.0016} & \textbf{0.0478} \\
\hline
MFCC + pBN   & 2.75 & 0.0130 & 0.4472 & \textbf{1.02} & \textbf{0.0038} & \textbf{0.1132} & \textbf{0.55} & \textbf{0.0016} & 0.0768 \\
\hline
MFCC + sBN   & 8.76 & 0.0550 & 0.9599 & 23.57 & 0.0981 & 0.9954 & 5.71 & 0.0348 & 0.9429 \\
\hline
\end{tabular}
\end{table*}

\begin{table}[htpb]
\caption{Comparison of different features on the male part of RedDots Part-4.}
\label{table:bn_part4}
\centering
\begin{tabular}{|c|c|c|c|}
\hline
Modeling & EER & MDCF08 & MDCF10 \\
\hline
\hline
MFCC        & \textbf{9.78} & \textbf{0.0463} & \textbf{0.9282}  \\
\hline
MFCC + pBN  & 15.04 & 0.0649 & 0.9903 \\
\hline
MFCC + sBN  & 15.87 & 0.0783 & 0.9703 \\
\hline
\end{tabular}
\end{table}

\section{Results}
\label{sec:results}

\subsection{Comparison of different modeling methods}
The performance of different systems on RedDots Part-1 is shown in Table \ref{table:result_part1}. We first look at the three error types. In this fixed-phrase condition, it is possible to detect the wrong text, even if no text recognition involved. Due to the mismatch between both speakers and phrases, all systems perform well on the Imposter-Wrong trials.

The first two lines show the results of conventional GMM-UBM and i-vector modeling. This confirms that the traditional i-vector framework performs worse than vanilla GMM-UBM in text-dependent speaker verification where the speech duration is very short. These methods have worse performance on the Target-Wrong trials since they are only designed to discriminate speakers and unable to capture the lexical information.

Then we go to the last four lines in Table \ref{table:result_part1}. As we see, HMM-based modeling achieves much better performance than the first two systems. The results improve the most on the Target-Wrong trials. The reason for this improvement is that the statistics of the HMM-based alignment contains more information about lexical contents, which is useful to reject wrong text. On the Imposter-Correct trials, HMM-based modeling methods also achieve greater than 20\% relative improvement (GMM-HMM to GMM-UBM, or i-vector/HMM to i-vector).

Interestingly, compared to GMM-UBM, we find that i-vector modeling makes better use of the HMM alignment. Even though the conventional i-vector works much worse than GMM modeling, when leveraged with HMM, i-vector/HMM achieves results on par with GMM-HMM. The GMM-HMM system outperforms i-vector/HMM on the Imposter-Correct trials, while i-vector/HMM performs the best on the wrong text trials.

The impact of the Imposter-Correct error on RedDots Part-4 are shown in Table \ref{table:result_part4}. Compared to the above fixed-phrase condition, the performance of all systems degrades in this prompted-phrase condition.
The degradation is partly due to the incomplete coverage of phonemes in the enrollment data. On the other hand, when phrases vary during enrollment and test, the diversity of pronunciation brings more intra-phoneme variability.

On this part, GMM-HMM and i-vector/HMM achieve similar results. Unexpectedly, the HMM-based systems do not exhibit an advantage, but has the equivalent performance with the GMM-UBM.
To explain this phenomenon, we should notice the mismatch between the HMM training and evaluation datasets. The training data of the HMM mainly comes from Librispeech clean part, with accents closer to US English \cite{librispeech}. In contrast, native and non-native speakers on RedDots are recruited worldwide. This accent mismatch causes HMM alignment errors on RedDots. On Part-1, the misalignments appear in the same manner in the enrollment and test utterances, and cause only minor issues. However, on Part-4, the systems with suboptimal HMM alignments make comparisons between different phonemes which bring negative influence to the verification results.

\subsection{Comparison of different HMM alignment algorithms}
The performance of the Viterbi and FB alignments is shown in the last four lines in Table \ref{table:result_part1} and Table \ref{table:result_part4}. On RedDots Part-1, though there is not much difference between the results of these two algorithms for GMM-HMM, the FB alignment brings more benefits to i-vector/HMM modeling. The EER achieves 10\%, 9\% and 9\% relative improvement under the three error conditions. On Part-4, i-vector/HMM with FB alignment is still slightly better than its Viterbi version.

\begin{figure}[htbp]
\centering
\subfloat[The posteriors of the second state of phoneme $\langle \text{n}\rangle$ and the difference between these two algorithms.]{\includegraphics[width=0.9\linewidth]{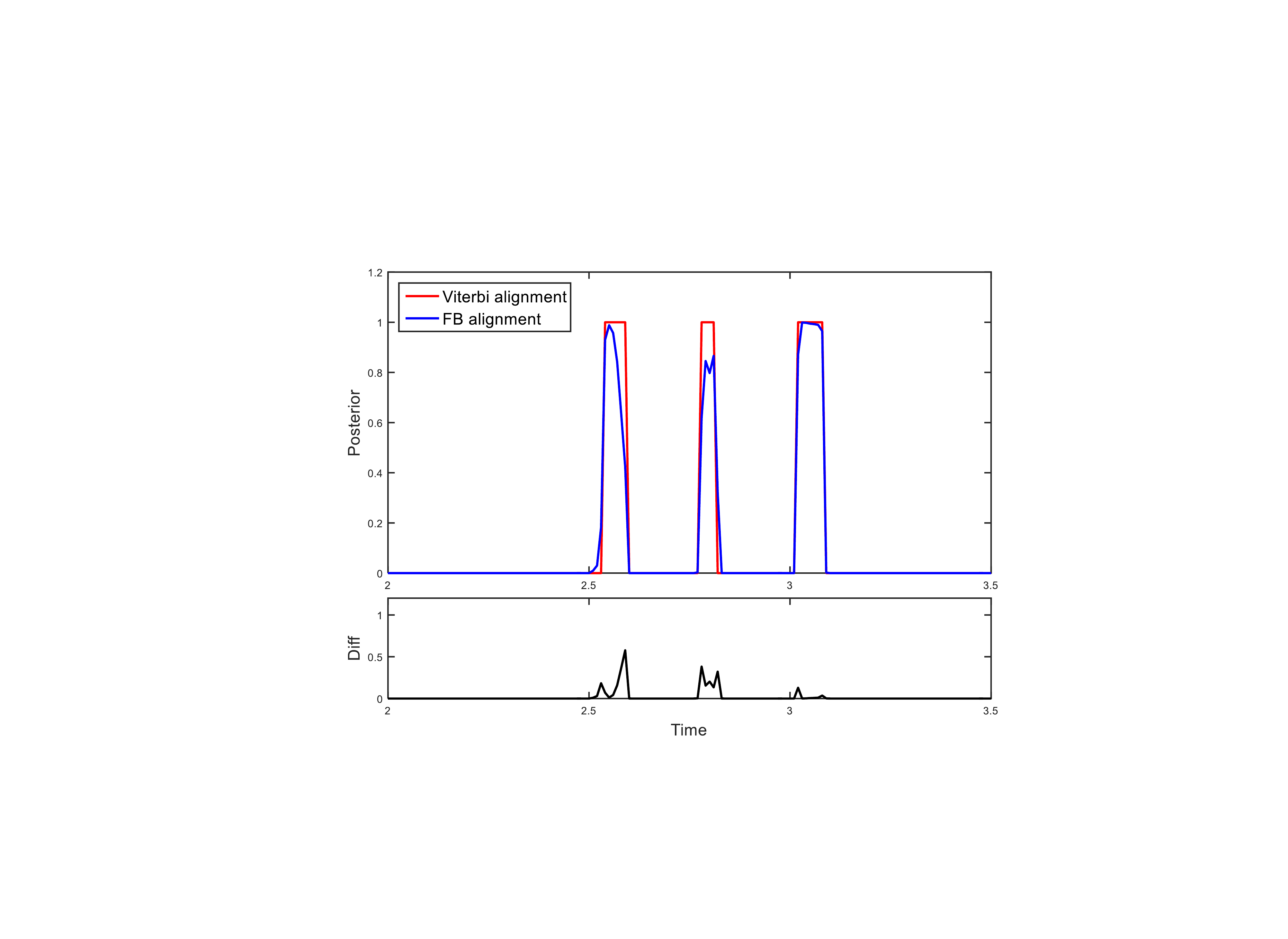}
\label{sfig:state_alignment}}
\hfil
\subfloat[The posteriors of one mixture in the above state. The results of the Viterbi and FB alignments are still similar to each other. ]{\includegraphics[width=0.9\linewidth]{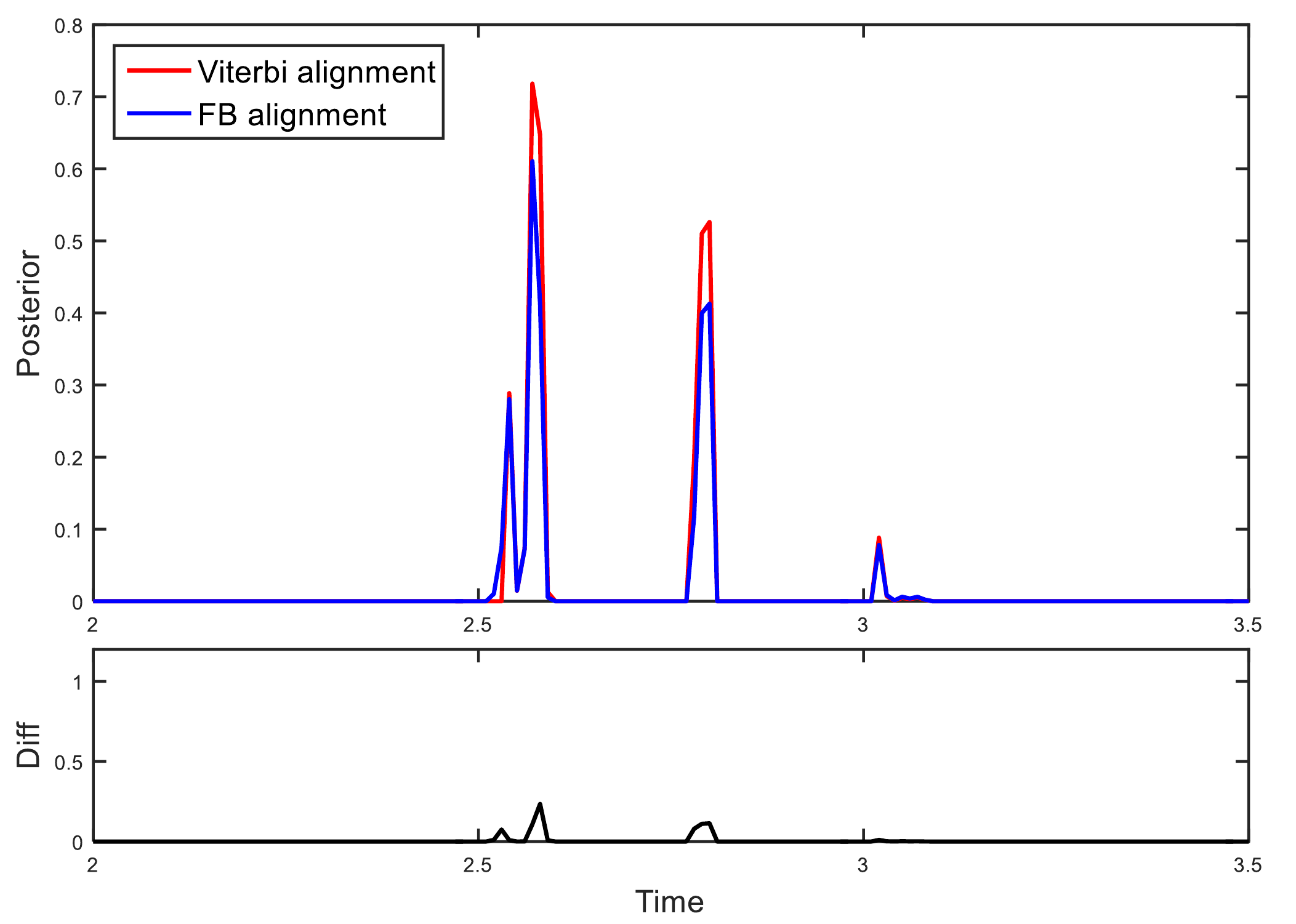}
\label{sfig:mix_alignment}}
\hfil
\caption{{The comparison of the Viterbi and FB algorithms. The utterance is randomly selected from the enrollment set of RedDots Part-1. }}
\label{fig:align_comparion}
\end{figure}

Although some prior work has suggested that the hard Viterbi alignment limits the effectiveness of the HMM alignment \cite{dnn_viterbi_ti}, it gives a satisfying performance in our experiments, especially for the GMM-HMM approach. It is worth mentioning that the Viterbi algorithm used in this paper aligns a frame to one phonetic state, rather than a single Gaussian senone. In each state, the alignment is again soft to mixture components. This hard-soft combination makes Viterbi alignment in our HMM-based systems relatively robust. We randomly choose one utterance in the enrollment set of RedDots Part-1 and illustrate the difference between the Viterbi and FB alignments in Fig. \ref{fig:align_comparion}.

From Fig. \ref{sfig:state_alignment}, we can see that in state-level, the output of Viterbi alignment is either 0 or 1 while the FB results range among $[0,1]$.
The Fig. \ref{sfig:mix_alignment} shows the alignment results of a single mixture. We can see that in both levels, the alignment results are very close to each other in most cases.
The i-vector framework may be more sensitive to frame alignments and benefits from the small difference.

\subsection{Comparison of different features}
We report the performance of different bottleneck features using the GMM-HMM modeling with Viterbi alignment. A traditional way to utilize bottleneck features is to concatenate it with MFCCs to from \emph{tandem features}. We use tandem features in the following experiments.

From Fig. \ref{table:bn_part1} and \ref{table:bn_part4}, we find that pBN improves the results on the wrong text trials because different phonemes are better discriminated by the phonetic information. In contrast, pBN performs worse for the Imposter-Correct trials due to the less speaker-dependent information. Considering our main target is to detect imposters, the characteristics of pBN are undesirable, because the wrong text can be relatively easy to be rejected by additional speech recognition systems.

Next, we analyze the performance of sBN. The results on the Target-Wrong trials confirm that the sBN training process eliminates the lexical diversity of one speaker. These tables also show that sBN does not achieve satisfactory performance in the Imposter conditions. We think this is because the dimension of sBN is too small to encode sufficient speaker information. In a typical d-vector system, the dimension of the BN feature is 400 to 600 \cite{deep_speaker_feature}, which is much larger than our setup. However, it is difficult to use the high-dimensional features in the above models.
Apart from utterance-level averaging, we need to develop a more sophisticated approach to fully utilize the information contained in sBN.

\section{Conclusions}
\label{sec:conclusion}
Four state-of-the-art modeling methods, including the GMM-UBM, i-vector, GMM-HMM, and i-vector/HMM approaches, are introduced in this paper. Their performance is compared under fixed- and prompted-phrase conditions on the RedDots dataset. In the fixed-phrase condition, the HMM-based modeling significantly outperforms conventional GMM-UBM and i-vector framework. In the prompted-phrase condition, the HMM alignment shows no advantages against GMM-UBM. This may be caused by the mismatch between the HMM training and evaluation data. Even though the i-vector system performs the worst, when coupled with HMM, the i-vector/HMM modeling performs comparable with GMM-HMM. 

Furthermore, we investigate the influence of the Viterbi alignment versus forward-backward alignment.
The posteriors calculated from the Viterbi and FB algorithms are close to each other. Even so, the i-vector/HMM modeling still benefits from the finer FB alignment. In practice, however, engineers would prefer the Viterbi algorithm because it has similar performance and lower computational complexity.

Different bottleneck features have been evaluated using the GMM-HMM modeling. Phonetic-dependent BN features perform well in the wrong text conditions, while they are still unable to achieve good results on the most important Imposter-Correct trials. On the other hand, speaker discriminant BN features work poorly in this model. This phenomenon may be due to the too small dimension of the bottleneck features.

In future work, we plan to improve the performance of HMM-based modeling in the prompted-phrase condition, as well as increase the speaker discriminative ability for the phonetic-dependent BN feature. And we will also study new models for the high-dimensional speaker discriminant BN feature.

\newpage
\bibliographystyle{IEEEbib}
\bibliography{mybib}

\end{document}